\documentclass[aps,twocolumn,showpacs,prl,amsmath,amssymb,floatfix,superscriptaddress]{revtex4-1}
\usepackage{amsmath,amssymb,graphicx,color}
\usepackage{bm}
\usepackage{bbm}
\usepackage[caption=false]{subfig}
\usepackage[usenames,dvipsnames]{xcolor}
\usepackage[colorlinks=true,citecolor=Cerulean,linkcolor=RubineRed,urlcolor=Cerulean]{hyperref}
\usepackage[normalem]{ulem}
\usepackage{soul,xcolor}

\newcommand{\mi}{\mathrm{i}} 

\newcommand{\Tr}{\ensuremath{\mathrm{Tr}\,}}
\newcommand{\avg}[1]{\ensuremath{\langle #1 \rangle}}

\usepackage{mathptmx} 

\begin{document}
\setstcolor{blue}
\title{Quantum Performance of Thermal Machines over Many Cycles}

\author{Gentaro Watanabe}
\affiliation{Department of Physics and Zhejiang Institute of Modern Physics, Zhejiang University, Hangzhou, Zhejiang 310027, China}
\affiliation{Center for Theoretical Physics of Complex Systems, Institute for Basic Science (IBS), Daejeon 34051, Korea}
\affiliation{University of Science and Technology (UST), 217 Gajeong-ro, Yuseong-gu, Daejeon 34113, Korea}
\author{B. Prasanna Venkatesh}
\affiliation{Institute for Quantum Optics and Quantum Information of the Austrian Academy of Sciences, Technikerstra\ss e 21a, Innsbruck 6020, Austria}
\affiliation{Institute for Theoretical Physics, University of Innsbruck, A-6020 Innsbruck, Austria}
\author{Peter Talkner}
\affiliation{Institut f\"{u}r Physik, Universit\"{a}t Augsburg, Universit\"{a}tsstra\ss e 1, D-86135 Augsburg, Germany}
\affiliation{Institute of Physics, University of Silesia, 40007 Katowice, Poland}
\author{Adolfo del Campo}
\affiliation{Department of Physics, University of Massachusetts, Boston, Massachusetts 02125, USA}
\begin{abstract}
The performance of quantum heat engines is generally based on the analysis of a single cycle. We challenge this approach by showing that the total work performed by a quantum engine need not be proportional to the number of cycles. Furthermore, optimizing the engine over multiple cycles leads to the identification of scenarios with a quantum enhancement. We demonstrate our findings with a quantum Otto engine based on a two-level system as the working substance that supplies power to an external oscillator.
\end{abstract}

\maketitle

Advances in technology have spurred the fabrication and study of thermal machines at the nanoscale, whose performance is governed by quantum fluctuations. 
Prominent examples include quantum heat engines (QHEs) and pumps \cite{scovil59,alicki79,Kosloff84,bender00}. Various prototypes have been realized in the laboratory by means of cold atoms and trapped ions as a working substance \cite{Brantut13,Rossnagel16}. Theoretical studies of these machines are largely motivated by foundational questions that address the interplay between thermodynamics and statistical mechanics in the quantum world \cite{kosloff14,Goold16}. At the same time, exciting applications are in view. Processes varying from laser emission \cite{scovil59} to light harvesting in both artificial and natural systems \cite{Scully11,Dorfman13,Killoran15} can be described in terms of QHEs.

Nonetheless, the quest for quantum signatures of the performance of thermal devices remains challenging. It is understood that a universal behavior emerges in the limit of small action \cite{Uzdin15}. Identifying scenarios exhibiting quantum supremacy, with a performance surpassing that in classical thermodynamics, stands out as an open problem. To this end, the use of quantum coherence \cite{Scully03}, nonequilibrium reservoirs \cite{Abah14,Rossnagel14}, and many-particle effects \cite{jaramillo15,Beau16} has been proposed.

The performance of quantum thermal machines is usually assessed via the characterization of a single cycle, as in classical thermodynamics. This approach assumes that the average single-cycle efficiency and power carry over to an arbitrary number of cycles, i.e., work done through $n$ cycles is expected to be equal to $n$ times the work done per cycle. 
Yet, in quantum mechanics work is determined via projective energy measurements at the beginning and end of a prescribed protocol \cite{Talkner07,Campisi11}.
As a result, assessing the performance of a quantum thermal machine can severely alter its dynamics due to the quantum measurement backaction. We argue that the QHE performance can be best assessed by measurements on an external system on which work is done (see, e.g., \cite{hayashi15} for a related discussion).
By analyzing the dynamics over many cycles, we elucidate the role of the intercycle coherence and find scenarios with quantum-enhanced performance.
In particular, we demonstrate that the average amount of work through $n$ cycles need not be proportional to $n$; rather, it may have an additional oscillatory contribution as a function of $n$.
Our work provides clear evidence that in the quantum regime the characterization of the QHE focused on a single cycle is insufficient. We propose that  assessments of the performance should address the global process over many cycles.

\begin{figure}[t!]
\centering
\includegraphics[width=0.72 \columnwidth]{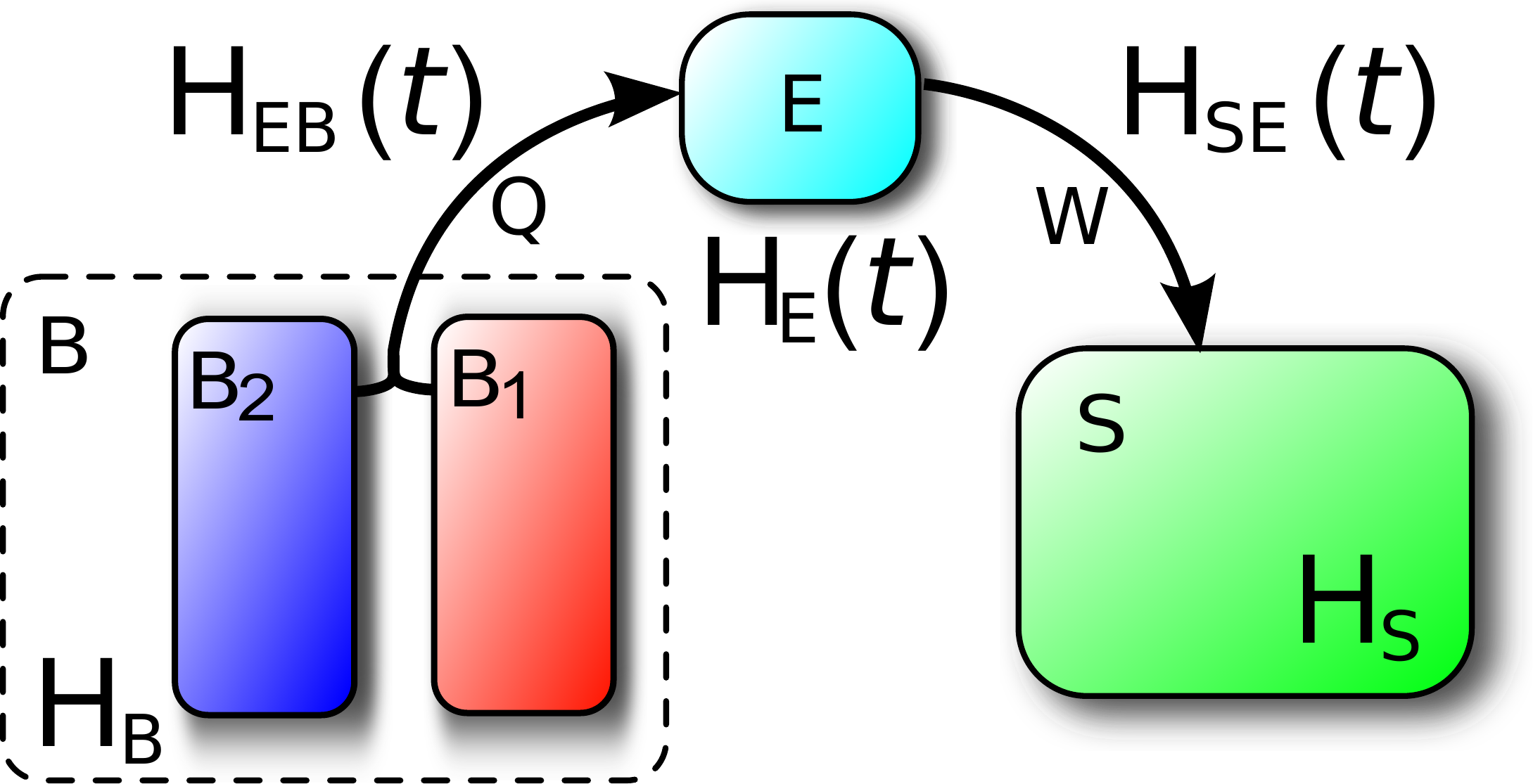}
\caption{
  {\bf Schematic quantum heat engine.}
  The quantum engine $E$ does work $w$ on an external system $S$ through the coupling $H_{SE}$ absorbing heat $Q$ from the baths collectively represented by $B$, which consists of hot ($B_1$) and cold ($B_2$) baths.}
\label{fig:setup}
\end{figure}

\textit{Setup.---}
We consider a quantum engine $E$ coupled to an external quantum system $S$ on which the engine does work (see Fig.~\ref{fig:setup}). The engine also interacts with heat baths $B$. The global Hamiltonian is the sum of that of the engine, the baths, the coupling between the engine and the baths, the system, and the coupling between the system and the engine:
\begin{equation}
  H(t) =  H_E(t) + H_B + H_{EB}(t) + H_S + H_{SE}(t)\, ,
\end{equation}
where the external system and the baths are assumed to be time independent. Under periodic driving over identical cycles, $H_E(t+T)=H_E(t)$, $H_{EB}(t+T)=H_{EB}(t)$, $H_{SE}(t+T)=H_{SE}(t)$, and  $H(t+T)=H(t)$, where $T$ is the period of one cycle. We further assume that the system-engine interaction $H_{SE}(t) = g_{SE}(t) \tilde{H}_{SE}$, where $g_{SE}(t)$ is a time-dependent coupling constant and $\tilde{H}_{SE}$ is a time-independent operator.

The work done by the engine is evaluated by energy measurements on the external system $S$. We consider two definitions of work. In the first one, the work $w$ done during $n$ cycles is evaluated by two energy measurements at the beginning and the end of $n$ cycles. In the second one, the work $\tilde{w}$ done over $n$ cycles is evaluated by $n+1$ energy measurements, one at the beginning at $t=0$ and one after the completion of each cycle. While in the classical case both definitions agree, this is no longer the case in the quantum regime, as we demonstrate next. For simplicity, we turn off the coupling $g_{SE}(t)$ at $t=0$, $T$, $\cdots$, $nT$; at these times, $[H_S, H(t)]=0$, and the energy eigenbasis of $H_S$, which is chosen to be the measurement basis, is shared by $H(t)$.
The external system is initially prepared in an energy eigenstate denoted by $|t=0\rangle_S=|0\rangle_S$ with eigenenergy $E_0^S$, i.e., $H_S|0\rangle_S = E_0^S |0\rangle_S$. The subindex $0$ here denotes $t=0$. The initial state $\rho_0$ of the total system reads $\rho_0 = \rho_0^{EB} \otimes |0\rangle_S{_S}\langle 0|$, where $\rho_0^{EB}$ is the initial state of the engine and bath parts.

\textit{Average of work over many cycles.---}
First, we consider the average of work $\avg{w}_{n}$ done on the system $S$ during $n$ cycles. Because of the periodicity of $H(t)$, the time evolution $U_{nT}$ of the total system from $t=0$ to $t=nT$ can be expressed as the $n$th power of the propagator $U_T =\mathcal{T} \exp {[- \mi \int_0^T dt H(t)]}$ ($\mathcal{T}$ is the time-ordering operator) of a single cycle, i.e., $U_{n T} = (U_T)^n$.
Thus, the average of work $\avg{w}_{n}$ is
\begin{align}
\avg{w}_{n}\! = \! \sum_i (E_i^S - E_0^S)\, \Tr\!_{EB}\!\left[{}_S\langle i| (U_T)^n \rho_0 (U_T^\dagger)^n |i\rangle_S\right],
\label{eq:w}
\end{align}
where $\Tr_{EB}[\cdots]$ denotes the trace over the Hilbert space of the engine and the baths, $|i\rangle_S$ the $i$th eigenvector of $H_S$, and $E_{i}^S$ the corresponding eigenvalue which is one of the possible results of an energy measurement.

To evaluate the second definition of work $\tilde w$, we perform energy measurements on the system $S$ at $t=T, 2T, \cdots, (n-1)T, nT$, where we obtain a result $k_1, k_2, \cdots, k_{n-1}, i$, respectively. Writing ${\bf k}\equiv (k_1, k_2, \cdots, k_{n-1})$ and summing over the intermediate states ${\bf k}$, the average of work $\avg{\tilde{w}}_{n}$ is given by
\begin{align}
  \avg{\tilde{w}}_{n} = \sum_i (E_i^S - E_0^S) \sum_{{\bf k}} \mathcal{T}_{i,0}^{{\bf k}; {\bf k}}\,
\label{eq:tildew}
\end{align}
with
\begin{align}
  \mathcal{T}_{i,0}^{{\bf k};{\bf k}'}
  &\equiv \Tr_{EB}\left[ \mathcal{M}_{i,k_{n-1}}^{EB} \cdots \mathcal{M}_{k_{2}, k_{1}}^{EB} \mathcal{M}_{k_{1},0}^{EB} \rho_0^{EB} \right. \nonumber\\
  & \quad \left. \times (\mathcal{M}_{k_{1}',0}^{EB})^\dagger (\mathcal{M}_{k_{2}', k_{1}'}^{EB})^\dagger \cdots (\mathcal{M}_{i,k_{n-1}'}^{EB})^\dagger \right]\, .
\label{eq:ti0}
\end{align}
where $\mathcal{M}_{i,j}^{EB} \equiv {}_S\langle i| U_T |j\rangle_S$ is the time evolution operator on the subspace spanned by  the engine and the baths. Writing $\avg{w}_{n}$ in terms of $\mathcal{T}_{i,0}^{{\bf k};{\bf k}'}$, we obtain
\begin{equation}
  \avg{w}_{n} = \sum_i (E_i^S - E_0^S) \sum_{{\bf k}, {\bf k}'} \mathcal{T}_{i,0}^{{\bf k}; {\bf k}'}\, ,
\end{equation}
where the sum over intermediate states runs over ${\bf k}$ and ${\bf k}'$.  By contrast, for $\avg{\tilde{w}}_{n}$ given by Eq.~(\ref{eq:tildew}), it runs only with respect to ${\bf k}$, as the intermediate measurements diagonalize the state, suppressing the intercycle quantum coherence in the system  $S$ on which work is done.

\textit{Model.---}
We next demonstrate that the average amount of work $w$ done over $n$ cycles is not proportional to $n$ in the quantum regime.
We choose a harmonic oscillator (HO) (with frequency $\omega$) as the external system $S$: $H_S = \omega a^\dagger a$
with $H_S|j\rangle_S = E_j^S |j\rangle_S = j\omega |j\rangle_S$. For simplicity, we initialize the external system $S$ in the ground state $j=0$ with $E_0^S=0$ at $t=0$.
Since the HO has an unbounded equidistant energy spectrum, energy can be deposited without an upper bound. We consider that, on the engine side, a two-level system (TLS) works as the interface with the external system, and the coupling is
\begin{equation}
  H_{SE}(t) = g_{SE}(t)\, \sigma_x\, (a^\dagger + a) . \label{eq:secouple1}
\end{equation}
Here, the Pauli matrix $\sigma_x\equiv \sigma_+ + \sigma_-$ with $\sigma_+ \equiv |e\rangle_E{_E}\langle g|$ and $\sigma_- \equiv |g\rangle_E{_E}\langle e|$ being the raising and lowering operators of the TLS, respectively, and $|g\rangle_E$ and $|e\rangle_E$ are the ground and excited states of the TLS, respectively.

\begin{figure*}[t]
  \centering
  \subfloat {\includegraphics[height=3.3cm]{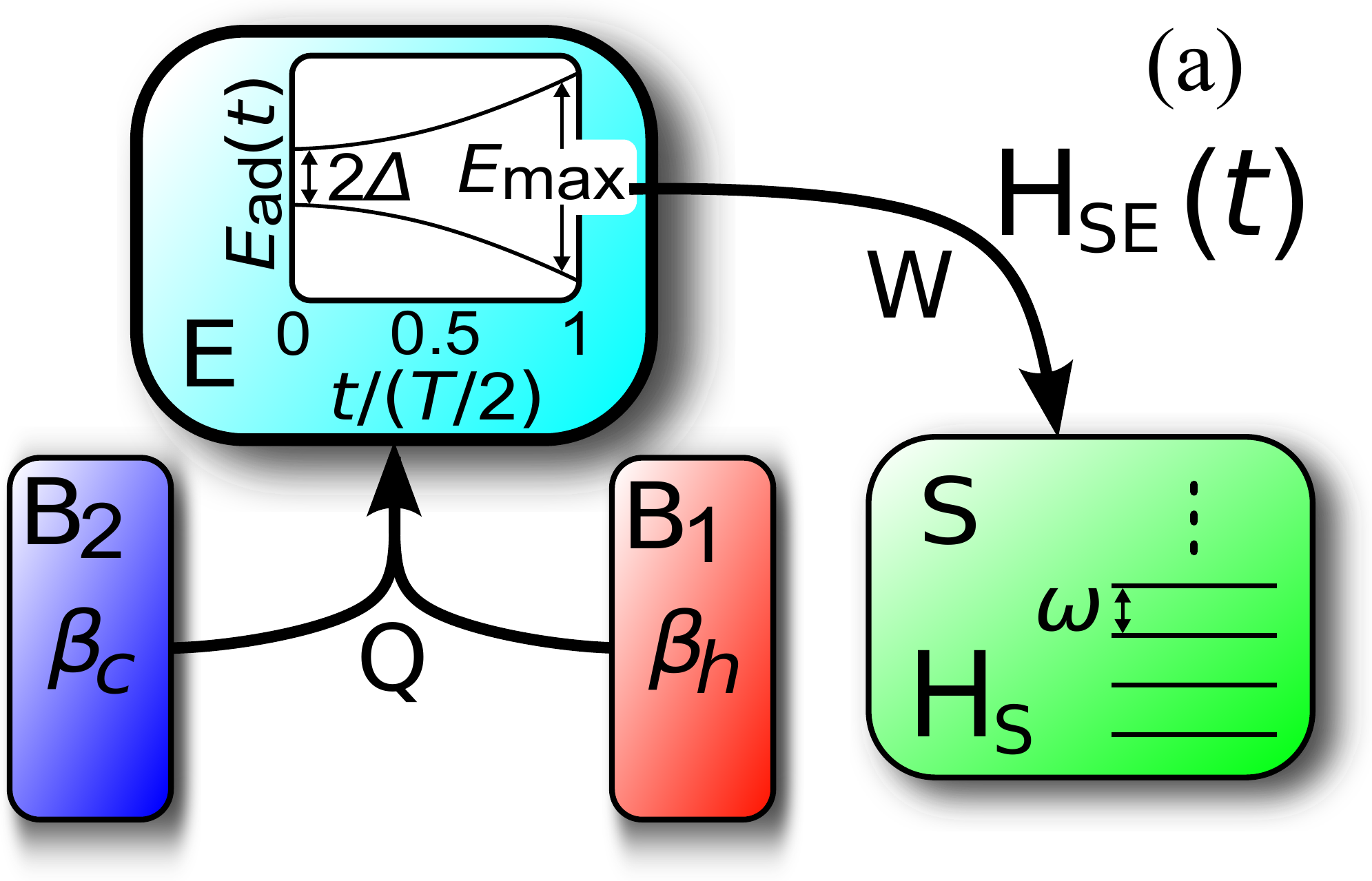}} \hspace{5mm} \subfloat {\includegraphics[height=3.3cm]{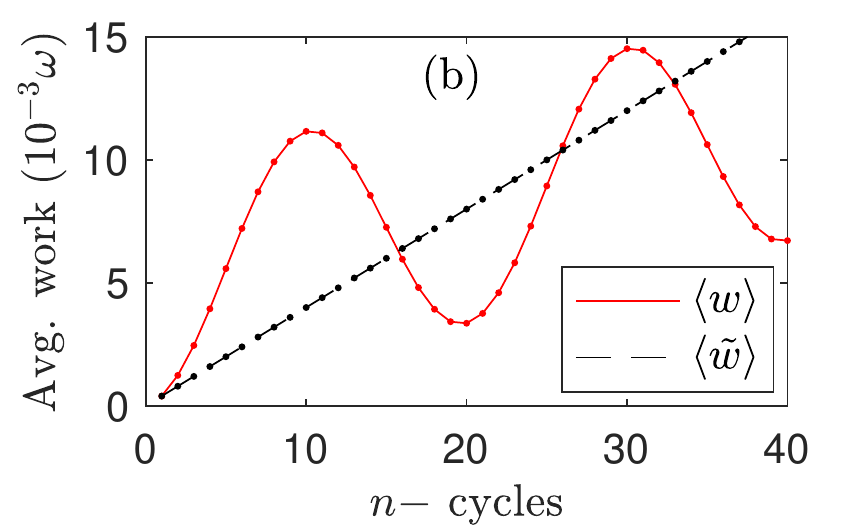}}\\
  \subfloat {\includegraphics[height=3.3cm]{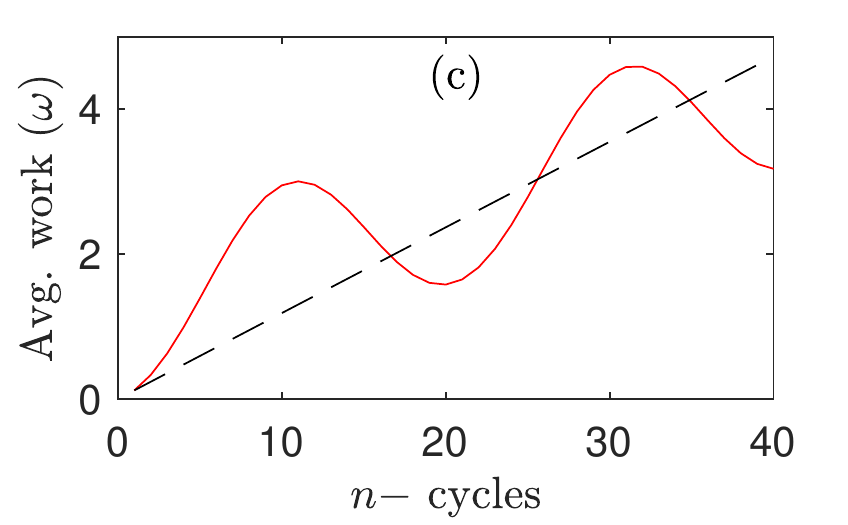}} \hspace{5mm} \subfloat {\includegraphics[height=3.3cm]{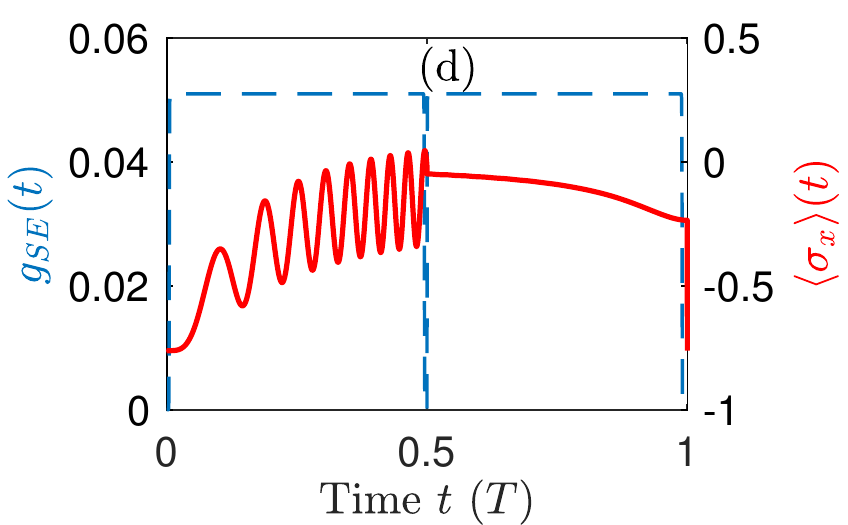}}
  \caption{
    {\bf Quantum performance of a heat engine.}
    (a) Schematic setup of a TLS engine ($E_{\rm ad}$ being the adiabatic energy levels) running with a hot ($B_1$) and cold ($B_2$) bath and coupled to a HO system $H_S$ via $H_{SE}$. Average work $\avg{w}_n$ and $\avg{\tilde{w}}_n$ as functions of the number of cycles $n$ for the perturbative ($g=0.02$) and impulse-type coupling [(b)] and for the nonperturbative ($g=0.5$, $\alpha T= 2142$) and continuous coupling [(c)]. Lines in (b) and (c) are from numerical calculations, and dots in (b) are from the analytical expressions (\ref{eq:wfinoffres}) and (\ref{eq:tildewn_fin}). (d) $g_{SE}(t)$ (cyan dashed line) and $\avg{\sigma_x^{(I)}(t)}_{\rho_0^{EB}}$ (red solid line) for the first cycle in the case of (c). The marked difference in the dynamics of $\sigma_x(t)$ in the two strokes of the engine comes from the interaction of the engine with the cold bath at $t=0$ and the hot one at $t=T/2$ leading to an almost pure state at $t=0$ and an almost mixed state at $t=T/2$ for the present choice of parameters. We set $b=0.1/\Delta$ in (b) and $\delta_t=0.98$ in (c) and (d). Other parameters are $\omega T = 0.05 \times 2\pi$, $v=0.5 \Delta^2$, $T = 20 /\Delta$, $\beta_h = 1/4 E_{\mathrm{max}}$, and $\beta_c = 1/\Delta$.}
\label{fig:otto}
\end{figure*}

We first consider an impulse-type coupling of the form
\begin{equation}
  g_{SE}(t) = g \sum_{m=0}^\infty \delta\left[t-\left(m + b \right)T\right]\, \label{eq:impulsecoupling}
\end{equation}
with a small coupling constant $g \ll 1$ and $0<b<1$ that allows a perturbative approach.
For this type of coupling, $\mathcal{M}_{i,j}^{EB}$ can be separated into the contributions from $H_E(t) + H_B + H_{EB}(t)$ and $H_{SE}(t)$ as
$\mathcal{M}_{i,j}^{EB} = U_{T,bT}^{EB}\, {}_S\langle i|e^{-\mi g\tilde{H}_{SE}}|j\rangle_S\, U_{bT,0}^{EB} e^{-\mi \omega [(1-b)i+ bj]T}$
with $U^{EB}_{t,0} \equiv \mathcal{T}\exp{[-\mi\int_0^{t}dt'\, H_E(t') + H_B + H_{EB}(t') ]}$.
On the rhs of Eq.~(\ref{eq:w}), contributions to the order of $g^2$ come from $i=0$ and $1$. Terms with $i\ge 2$ contribute only to $O(g^4)$ or higher. In addition, only terms with $i=1$ give nonzero values of work $w$. Thus, we obtain to leading order
\begin{align}
  \avg{w}_{n}
  &\simeq \omega g^2 \sum_{m, m' = 0}^{n-1} e^{\mi\omega (m-m')T}\,\nonumber\\
  & \quad \times \avg{ \sigma_x^{(I)}\left[(m'+b)T\right] \sigma_x^{(I)}\left[(m+b)T \right] }_{\rho_0^{EB}}\, ,
\label{eq:w_fin}
\end{align}
where $\sigma_x^{(I)}(t) \equiv {U^{EB}_{t,0}}^\dagger \sigma_x U^{EB}_{t,0}$ is the operator $\sigma_x$ in the interaction picture and $\avg{\cdots}_{\rho_0^{EB}} \equiv \Tr_{EB} \left[ \cdots \rho_0^{EB} \right]$.

The rhs is determined by two-time correlation functions of the engine operator $\sigma_x^{(I)}$ at different multiples of the cycle period $T$. At equal times, $m=m'$, the correlation functions become equal to one.  We assume that the  working substance of the heat engine  undergoes  complete thermalization within each cycle; therefore,  the correlation functions at different times are factorized to be $\avg{\sigma_x^{(I)}\left[(m'+b)T\right] \sigma_x^{(I)}\left[(m+b)T \right] }_{\rho_0^{EB}} = \avg{ \sigma_x^{(I)}(bT) }_{\rho_0^{EB}}^2$. Finally the average of work becomes
\begin{align}
  \avg{w}_{n} &\simeq \omega g^2 \bigg\{ \avg{\sigma_x^{(I)}(bT)}_{\rho_0^{EB}}^2 \frac{\cos{(n \omega T)}-1}{\cos{(\omega T)}-1} \nonumber\\
    & \quad + \left[ 1-\avg{\sigma_x^{(I)}(bT)}_{\rho_0^{EB}}^2 \right] n \bigg\}\, ,
\label{eq:wfinoffres}
\end{align}
which presents a nontrivial dependence on $n$: An oscillatory $\cos(n \omega T)$ contribution is superimposed on the expected term proportional to $n$.
The interplay between these oscillatory and linear terms in $\avg{w}_n$ is a signature of quantum engines. When the HO becomes resonant with the engine cycle, i.e., for $\omega T = 2 \pi r$ with an integer $r$, the oscillatory term turns into a steady increase of the work proportional to $n^2$, because $\lim_{x \to r} [\cos(2 \pi nx) -1]/[\cos(2 \pi x)-1] =n^2$. Since this is due to the continuous injection of the energy from the time-dependent coupling constant instead of the engine, we will avoid the resonance point in the later discussion.

Also for $\avg{\tilde{w}}_{n}$, nonzero contributions of the order of $g^2$ come only from  $i=0$ and $1$. From Eq.~(\ref{eq:ti0}), one finds $\sum_{\bf k} \mathcal{T}_{1,0}^{{\bf k};{\bf k}} \simeq 1- \sum_{\bf k} \mathcal{T}_{0,0}^{{\bf k};{\bf k}} \simeq ng^2$, and hence from Eq.~(\ref{eq:tildew}) one obtains
\begin{equation}
  \avg{\tilde{w}}_{n} \simeq n\omega g^2\, ,
\label{eq:tildewn_fin}
\end{equation}
which is strictly proportional to $n$.
Regarding the higher moments of $w$ and $\tilde{w}$, $\avg{w^m}_n$ and $\avg{\tilde{w}^m}_n$ are given by $\simeq \omega^m p_n(1)$ with the probability $p_n(1)$ to obtain the final state $i=1$ after $n$ cycles. Therefore, $w/\omega$ and $\tilde{w}/\omega$ follow a Poisson distribution with the parameter $\lambda = p_n(1)$ to leading order with respect to the coupling constant $g$.

\textit{Numerical results for an Otto cycle.---} Our conclusions hold for realistic smooth functions $g_{SE}(t)$ with a wide range of values of the coupling strength $g$, governing the interaction between the engine and the system  during each work stroke. For the sake of illustration, we choose $g_{SE}(t)$ with the form
\begin{align}
  g_{SE}(t) &= \frac{g}{\delta_t T} \sum_{n=0}^\infty \left\{ \tanh \left[\alpha \left(t-t_1-\tfrac{n T}{2}\right)\right]\right.\nonumber\\
& \quad \left. - \tanh \left[\alpha \left(t-t_2-\tfrac{n T}{2}\right) \right] \right\}
\label{eq:squarecoupling}
\end{align}
with a fast switching rate $\alpha$. This coupling function takes nonzero values in the interval between $t_1$ and $t_2 = t_1 + \delta_t T/2$ with $0 < \delta_t < 1$ and vanishes approximately in the remaining part.

We numerically study the performance of a heat engine in a quantum Otto cycle \cite{quan07,Abah12} using a TLS as a working substance; see Fig.~\ref{fig:otto}(a).  The dynamics includes the  initialization and repetition of the four strokes of the cycle:

(0) \emph{Initial state.---} With  $g_{SE}(0) = 0$,  the TLS with Hamiltonian $H_E(0) = \Delta \sigma_x$ is prepared in thermal equilibrium with the cold bath at  inverse temperature $\beta_c$, from which it is decoupled at $t=0$.
The reduced density operator for the engine and the external system is $\rho(0) = Z_{\beta_c}^{-1}(0) \exp \left[-\beta_c H_E(0)\right] \otimes \vert 0 \rangle_S{_S}\langle 0 \vert$ with the partition function $Z_{\beta_c}(0)=\Tr_{E}\exp{\left[-\beta_c H_E(0)\right]}$, where $\Tr_{E}$ denotes a trace over the engine degrees of freedom. The energy separation of the TLS at the initial time is $2 \Delta$.

(1) \emph{Isentropic compression.---} From $0\leq t < T/2$, the engine remains decoupled from the heat baths and the total Hamiltonian changes according to $H_1(t) = H_{E}(t) + H_S + H_{SE}(t)$ with $H_E(t) = \Delta \sigma_x - vt \sigma_z$, where $v$ is a linear sweep rate.
The state of the engine and system at the end of the stroke is $\rho(T^-/2) = U_1 \rho(0) U_1^{\dagger}$ with $U_1 = \mathcal{T} \exp{[ -\mi \int_0^{T/2} dt\, H_1(t)]}$ (where $T^- \equiv T -\epsilon$ with an infinitesimal positive $\epsilon$). At $t=T/2$, the TLS energy separation takes its maximum value of $E_{\mathrm{max}} = 2 \sqrt{\Delta^2+(vT)^2/4}$.

(2) \emph{Hot isochore.---} At $t=T/2$, setting $g_{SE}=0$, the TLS thermalizes with the hot bath at inverse temperature $\beta_h$  in a negligible time \cite{note:thermalization}. 
At the end of the stroke, the reduced density operator is given by $\rho(T/2) = Z_{\beta_h}^{-1}(T/2)\exp{[-\beta_h H_E(T/2)]} \otimes \Tr_{E} \rho(T^{-}/2)$.

(3) \emph{Isentropic expansion.---} In the interval $T/2 \leq t < T$, the engine remains decoupled from the baths and evolves unitarily according to the Hamiltonian $H_2(t) = H_E(t) + H_S + H_{SE}(t)$ with $H_E(t) = \Delta \sigma_x + v (t-T) \sigma_z$. At the end of the stroke, the density matrix reads $\rho(T^{-}) = U_2 \rho(T/2) U_2^{\dagger}$ with $U_2 = \mathcal{T} \exp{[ -\mi \int_{T/2}^{T} dt\, H_2(t)]}$.

(4) \emph{Cold isochore.---} At $t=T$, setting $g_{SE}=0$, the TLS is brought into contact with the cold bath and quickly thermalizes such that the engine returns to the initial state, $\rho(T) = Z_{\beta_c}^{-1}(0) \exp{\left[-\beta_c H_E(0)\right]} \otimes \Tr_{E}\rho(T^{-})$. This is taken as the initial state for any new cycle, starting with stroke 1 (i.e., isentropic compression).

\begin{figure}[t]
\centering
  {\label{fig:fig3a}\includegraphics[width=0.65\columnwidth]{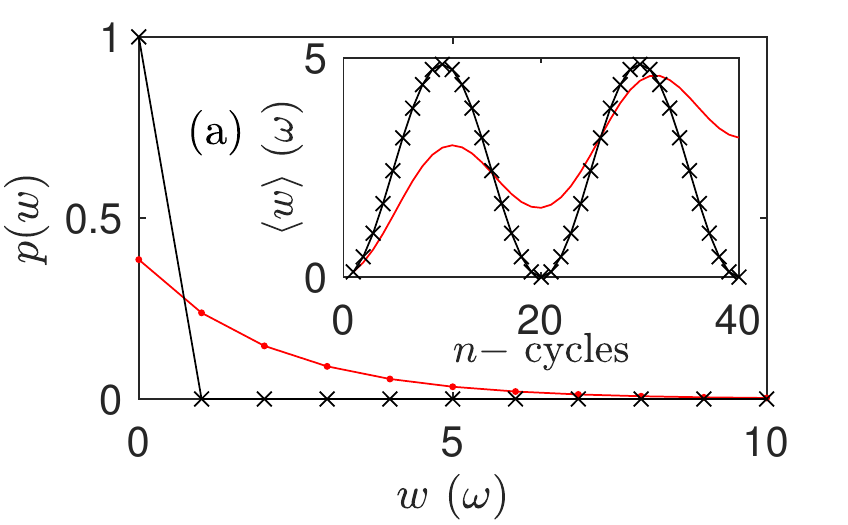}}
  \hfill
  {\label{fig:fig3b}\includegraphics[width=0.65\columnwidth]{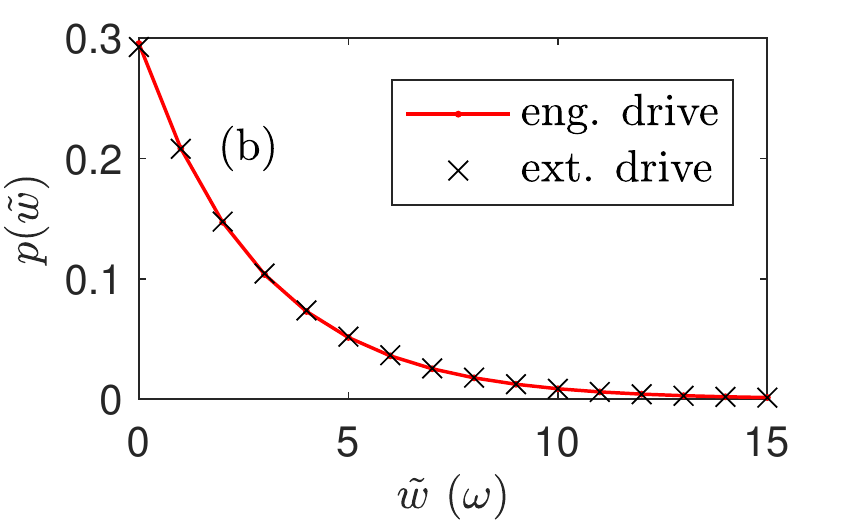}}
  \caption{
    {\bf Quantum work statistics.}
    Comparison of the probability distribution functions $p$'s of work (a) $w$ and (b) $\tilde{w}$ obtained in our numerical calculations for the quantum engine shown in Fig.~\ref{fig:otto}(c) and those for a classical force $f$ in Eq.~(\ref{eq:drivenho}). Here, $p$'s after 20 cycles are shown. The inset in (a) compares $\avg{w}_n$ performed by a quantum engine and by a classical force.
}
\label{fig:workstat}
\end{figure}

First, we consider the impulse-type coupling given by Eq.~(\ref{eq:impulsecoupling}) and compare the analytical expressions (\ref{eq:wfinoffres}) and (\ref{eq:tildewn_fin}) with the numerical results in the perturbative regime of $g \ll 1$. Figure \ref{fig:otto}(b) presents an excellent agreement between the analytic and the numerical results  for both $\avg{\tilde{w}}_n$ and $\avg{w}_n$. 
The effect of the oscillation of $\avg{w}_n$ is most important for small cycle number  $n$ when the oscillation amplitude of $\avg{w}_n$ is comparable to the linear component. Remarkably, $\avg{w}_n$ can surpass $\avg{\tilde{w}}_n$ for small $n$ ($n \le 15$ and $26 \le n \le 32$ in this example). This enhancement of the work is a consequence of the intercycle quantum coherence of the system.
By repeatedly performing the energy measurements at intervals of an optimum number of cycles (e.g., around every ten cycles for this case), we obtain linear scaling with respect to this interval of cycles but with a much larger slope. On the other hand, if the performance of the engine is evaluated by the work $\avg{w}_1 = \avg{\tilde{w}}_1$ extracted only through a single cycle, the slope of the linear scaling is overestimated as $\omega g^2$, while the true asymptotic value of the slope is $\omega g^2 \big[1-\avg{\sigma_x^{(I)}(bT)}_{\rho^{EB}_0}^2\big]$.

Figure \ref{fig:otto}(c) shows the numerical results for a nonimpulse square-type coupling $g_{SE}(t)$ given by Eq.~(\ref{eq:squarecoupling}) with a finite duration $\delta_t T/2$ with $\delta_t = 0.98$ from $t_1=0.005 T$ to $t_2 = 0.495 T$ and from $t_1 = 0.505 T$ to $t_2 = 0.995 T$ in each cycle [blue dashed line in Fig.~\ref{fig:otto}(d)].
We observe that the oscillation of $\avg{w}_n$ persists in spite of the fact that $\avg{\sigma_x^{(I)}(t)}_{\rho_0^{EB}}$ oscillates during the time in which the system interacts with the engine, i.e., when $g_{SE}(t) \neq 0$ as shown in the first half cycle in Fig.~\ref{fig:otto}(d). This confirms that the oscillatory dependence of $\avg{w}_n$ on $n$ is not an artifact of the impulsive coupling but rather a generic feature.

Finally, we pose the question whether the quantum nature of the engine does play a role. For this purpose, we replace the engine by a time-periodic classical force:
\begin{align}
  H(t) = \omega a^\dagger a - f(t) (a^\dagger + a),
  \label{eq:drivenho}
\end{align}
where the force $f(t)=f(t+T)$ has the period $T$ of the engine cycle.
Starting from the ground state of the HO, one may determine the full statistics of work \cite{talkner08}. It turns out that the details of the time dependence of the force within one period are irrelevant; only the magnitude of $|\int_0^T dt\, f(t) e^{\mi \omega t}|$  matters. We set this parameter in such a way that $\langle w \rangle_1$ performed in a single period is equal to the one delivered by the engine.
In Fig.~\ref{fig:workstat}(a), we compare the probability distribution function (PDF) $p(w)$ to obtain the work $w$ for the classical force and that for the engine. The two distributions pronouncedly differ from each other.
The difference is also apparent for the average work as a function of the number of cycles; see the inset in Fig.~\ref{fig:workstat}(a). It oscillates periodically and remains bounded for the classical force in contrast to the one for the engine with an overall linear increase.

The situation is totally different for $\tilde{w}$. As shown in Fig.~\ref{fig:workstat}(b), its PDF $p(\tilde{w})$ for the engine is very well reproduced by the classical force.
Therefore, with respect to $\tilde{w}$, the effect of the engine on the external system is trivial, in the sense that it can be reproduced by a classical driving. Effects of the quantum engine which cannot be mimicked by a classical force can be observed in $w$, while they are absent in $\tilde{w}$.

Our work demonstrates that the characterization of a quantum thermal machine based on its performance for a single cycle does not carry over multiple cycles, as it neglects the quantum coherence of the external system on which work is done. In particular, the work done over many cycles need not be directly proportional to the value measured over a single cycle and can exhibit an oscillatory behavior with respect to the number of cycles. 
By performing stroboscopic energy measurements at intervals of an optimum number of cycles, work can be extracted at a quantum-enhanced rate.  In addition,  while the full statistics of work measured over a single cycle can be reproduced by a classical external force, this is no longer the case when the performance of a quantum engine is assessed over multiple cycles. Our results should find broad applications in the design of energy-efficient thermal machines at the nanoscale.

\begin{acknowledgments} 
It is a pleasure to thank Mathieu Beau for useful comments on the manuscript. 
B.~P.~V. is funded by the Austrian Federal Ministry of Science, Research, and Economy (BMWFW), and he thanks Professor Oriol Romero-Isart for support.
P.~T. thanks the Polish Foundation for Science (FNR) for granting him an Alexander von Humboldt Honorary Research Fellowship.
This work was supported by the Zhejiang University 100 Plan, by the Junior 1000 Talents Plan of China, by IBS of Korea through Project Code (IBS-R024-D1), by NSF of China (Grant No. 11674283), by UMass Boston (P20150000029279), and by the John Templeton Foundation.
\end{acknowledgments}


\begin{thebibliography}{99}
\bibitem{note:thermalization} Here, the instantaneous equilibration is assumed just for simplicity. For more realistic discussions, one should note that the fast equilibration requires strong coupling with thermal baths, which leads to a deviation from the canonical distribution.
\end{thebibliography}
\end{document}